\documentclass[a4paper
               ]{jacow}
\usepackage{stfloats}
\usepackage{pdfpages,multirow,ragged2e} %

\makeatletter%
	\ifboolexpr{bool{xetex}}
	 {\renewcommand{\Gin@extensions}{.pdf,%
.png,.jpg,.bmp,.pict,.tif,.psd,.mac,.sga,.tga,.gif,%
	                    .eps,.ps,%
	                    }}

\usepackage{tikz}
\usetikzlibrary{arrows.meta, positioning}
\makeatother

\ifboolexpr{bool{xetex} or bool{luatex}} 
 {}                                      
 {\usepackage[utf8]{inputenc}}           

\usepackage[USenglish]{babel}

\begin{document}

\title{Integration of Retrieval-Augmented Generation for Knowledge Access in the ELBE Accelerator Control System}

\author{N. Mirian\thanks{n.mirian@hzdr.de} 
\\  Helmholtz-Zentrum Dresden-Rossendorf HZDR, Dresden Germany}
	
\maketitle

\begin{abstract}
The efficient operation of accelerator facilities increasingly relies on rapid access to heterogeneous operational knowledge, including logbooks, interlock reports, machine parameters, and historical archive data. At ELBE, we proposed a Retrieval-Augmented Generation (RAG) framework that integrates facility documentation and operational records into a unified AI-assisted support tool for operators. The system is expected to index electronic logbooks, machine archive time-series data, and subsystem manuals using domain-adapted embeddings stored in a vector database. User queries will be expected to be processed through a large language model that retrieves the most relevant operational context and generates structured, operator-oriented responses with traceable source references. 
This contribution presents the system architecture, data integration strategy, and challenges toward real-time AI-assisted accelerator operation.

\end{abstract}

\section{INTRODUCTION}

Modern accelerator facilities operate with increasing technical complexity and stringent availability requirements. Reliable beam delivery depends on the coordinated performance of radio-frequency systems, magnets, cryogenics, diagnostics, vacuum, and interlock subsystems, each producing large volumes of operational data. Machine states are documented in electronic logbooks, archived time-series databases, subsystem reports, and technical manuals. While essential for troubleshooting and performance optimization, this information is typically distributed across heterogeneous and weakly connected data sources.

At the electron linear accelerator ELBE (Electron Linac for beams with high Brilliance and low Emittance) at the Helmholtz-Zentrum Dresden-Rossendorf (HZDR) \cite{ELBE}, operators frequently analyze historical machine states to diagnose beam interruptions, interlock trips, RF instabilities, or performance drifts. This process often requires manual inspection of logbook entries, correlation of archived machine parameters, and consultation of subsystem documentation. As a result, troubleshooting can be time-consuming and highly dependent on operator experience, particularly when similar fault patterns occurred in the past but are difficult to retrieve through conventional keyword-based searches.

The growing volume of archived operational data makes systematic knowledge reuse increasingly challenging. Traditional database queries efficiently access numerical time-series data but do not support semantic search across textual documentation. Conversely, large language models alone lack reliable access to facility-specific operational records. A method that combines semantic retrieval of structured and unstructured data with controlled language generation is therefore desirable.

In this work, we propose a Retrieval-Augmented Generation (RAG) \cite{RAG} framework for operational support at ELBE. The system will integrate electronic logbooks, archived machine parameters, interlock reports, and subsystem manuals into a unified retrieval pipeline based on domain-adapted embeddings and a vector database. User queries are intended to process by a large language model that retrieves relevant historical context and will generate structured responses with source traceability. The goal is to reduce troubleshooting time and improve access to accumulated operational knowledge.

\section{OPERATIONAL CONTROL SYSTEM AND DATA ENVIRONMENT AT ELBE}

The electron linear accelerator ELBE operates in continuous-wave (CW) mode as a multi-user facility providing electron beams and secondary radiation for scientific applications. Commissioned in 2001, the facility consists of a superconducting RF linac \cite{LINAC}, an injector system, magnetic beam transport lines, undulator sections, and multiple experimental beamlines. Reliable operation requires continuous monitoring and coordination of RF systems, magnet power supplies, cryogenic infrastructure, vacuum systems, diagnostics, and machine protection components. 

\begin{figure*}
    \centering
    \includegraphics[width=11 cm, height=8 cm]{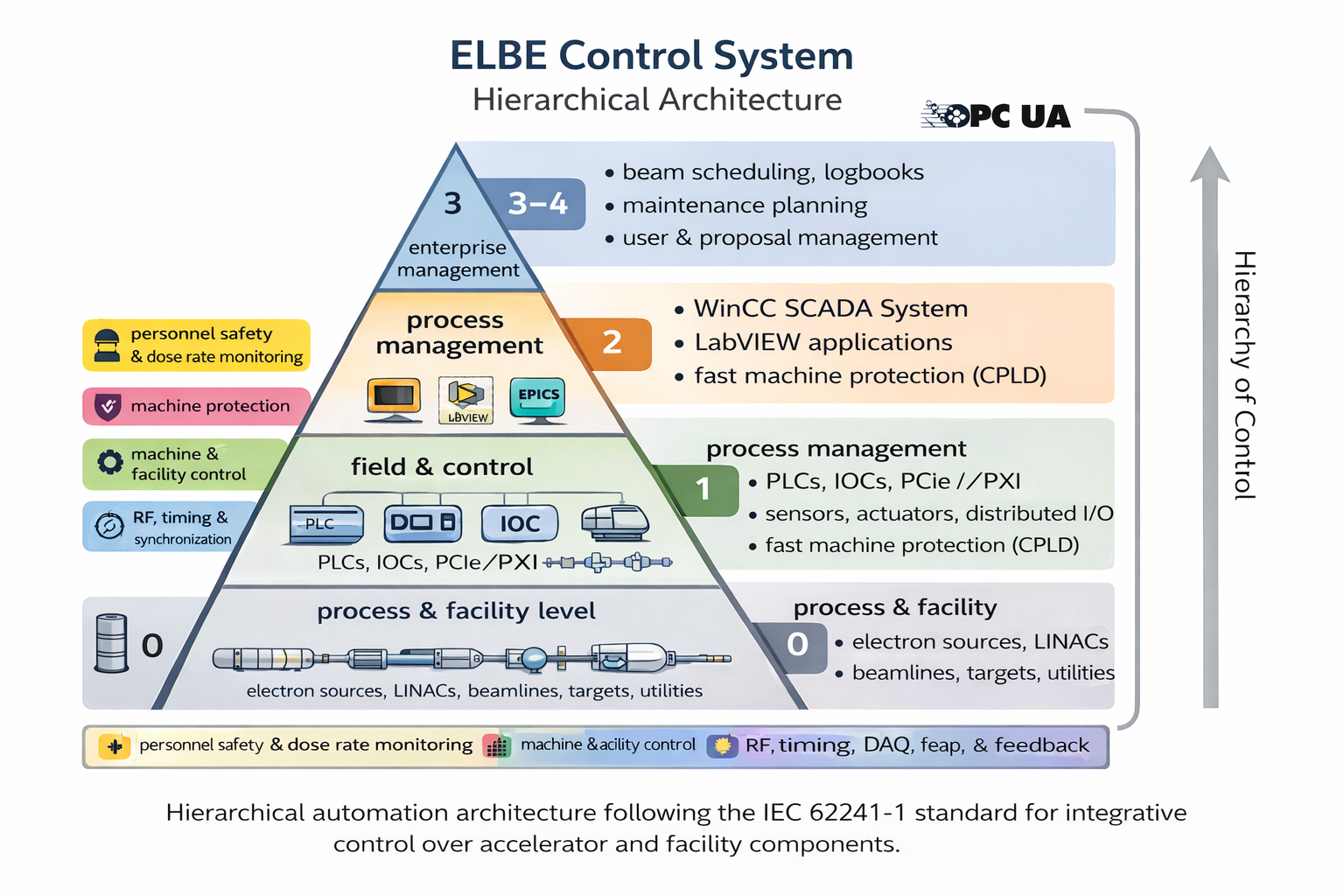}
    \caption{Hierarchical architecture of the ELBE control system based on the IEC 62241-1 automation model. The figure was generated using ChatGPT \cite{CGPT}}.
    \label{fig:elbe}
\end{figure*}

The ELBE control system follows a hierarchical industrial automation architecture \cite{Poster} based on the IEC 62241-1 standard \cite{IEC62264}. It integrates accelerator subsystems using several industrial control technologies, with ongoing developments focusing on improved device integration and data accessibility via OPC UA \cite{SIMATIC, OPC}. The architecture is organized into several layers (see Fig.~\ref{fig:elbe}). The process and facility level (Level 0) comprises the physical accelerator infrastructure including electron sources, LINACs, beamlines, targets, and utility systems. The field and control level (Level 1) contains sensors, actuators, front-end electronics, PLCs, IOCs, and distributed I/O systems responsible for machine control, diagnostics, and interlocks, including a fast machine protection system based on CPLD hardware logic. The process management level (Level 2) provides operator interfaces and high-level applications such as WinCC SCADA \cite{SIMATIC}, LabVIEW diagnostic tools \cite{NI}, and EPICS GUIs \cite{EPICS} for beamline, LLRF, and timing control. At the top, the enterprise management level (Levels 3–4) supports operational management functions including beam scheduling, electronic logbooks, maintenance planning, and user management. Operational data at ELBE can be broadly categorized into three classes:
\textbf{Structured time-series data}, consisting of archived machine parameters stored in historical databases with subsystem-dependent sampling rates;
\textbf{event-based records}, including interlock triggers, subsystem warnings, and machine state transitions; and
\textbf{unstructured textual documentation}, primarily electronic logbook entries describing machine operation, anomalies, tuning procedures, and recovery actions.
Although these data sources are individually accessible, they are not semantically linked. Correlating beam interruptions with archived parameter variations and relevant logbook entries typically requires manual cross-referencing across multiple systems. This lack of unified semantic access represents a significant operational bottleneck, particularly when similar fault patterns occurred previously but are difficult to identify. These heterogeneous data sources form the basis for the RAG framework described in the following sections.

 \section{RAG IMPLEMENTATION AND DATA INTEGRATION}


The proposed framework aims to reduce troubleshooting time and improve access to accumulated operational knowledge by enabling unified natural-language access to heterogeneous operational knowledge. Key requirements include compatibility with structured time-series data and unstructured documentation, robustness to accelerator-specific terminology and abbreviations, traceability of generated responses to original sources, and low-latency interaction suitable for operator workflows. The system is expected to be designed as a decision-support tool and does not perform control actions.

\subsection{Data Ingestion and Normalization}

The ingestion pipeline is expected to integrate electronic logbooks, interlock and machine protection records, archived process variables, and subsystem documentation. Textual sources (logbooks and manuals) would be converted to a common format and segmented into semantically coherent chunks. Each chunk would be enriched with metadata such as time stamps, subsystem labels (RF, magnets, vacuum, cryogenics, diagnostics, machine protection), component identifiers, and source type.
Event-based records are intended to be stored as structured entries containing event time, subsystem, and affected components, with links to relevant documentation where available. Time-series archive data would be retrieved within user-defined time windows. Parameter names would be normalized and associated with subsystem metadata, and selected archive segments would be summarized (e.g.\ statistics or trends) to provide compact contextual information for language model queries.

\begin{figure*}[htb]
\centering
\begin{tikzpicture}[
block/.style={draw, rectangle, rounded corners, align=center, minimum height=1cm, minimum width=2.8cm},
arrow/.style={-Latex, thick}
]

\node[block] (data) {Data Sources\\
Logbook\\Archive\\Interlocks\\Manuals};

\node[block, right=1.8cm of data] (index) {Indexing Layer\\
Chunking\\Metadata\\Embeddings\\Vector DB};

\node[block, right=1.8cm of index] (retrieve) {Retrieval Layer\\
Similarity Search\\Metadata Filtering\\Time Alignment};

\node[block, right=1.8cm of retrieve] (llm) {Generation Layer\\
LLM\\Structured Output\\Source Traceability};

\draw[arrow] (data) -- (index);
\draw[arrow] (index) -- (retrieve);
\draw[arrow] (retrieve) -- (llm);

\end{tikzpicture}
\caption{Layered architecture of the RAG-based operational support framework. Heterogeneous accelerator data are normalized and embedded into a vector database. User queries trigger semantic retrieval with metadata and time constraints before structured response generation.}
\label{fig:rag_architecture}
\end{figure*}
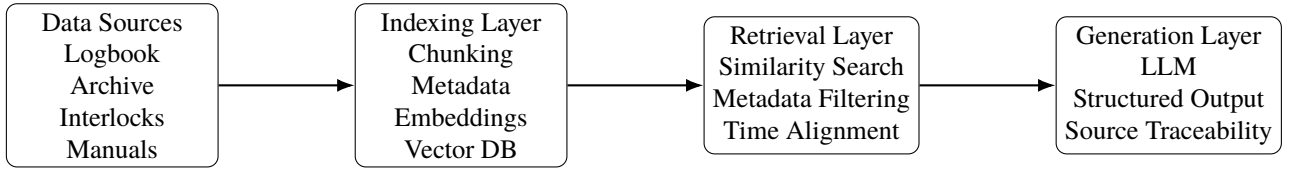

\subsection{Embedding and Retrieval}

Text chunks and event records are intended to be embedded using a domain-adapted embedding model and indexed in a vector database together with metadata.
This would enable hybrid retrieval combining semantic similarity search with metadata constraints such as subsystem filters or time ranges.
When a query is submitted, the system would perform query normalization (including acronym and component-name handling), retrieve the top-$k$ candidate contexts, and optionally re-rank them based on subsystem relevance, time proximity, and source reliability.
The retrieved contexts are intended to be assembled into a structured prompt for the language model.

\subsection{Response Generation and Traceability}

The large language model \cite{LLM, LLMSurvey} is expected to operate strictly in a retrieval-augmented mode, generating responses conditioned on the retrieved contexts. Outputs would be structured for operator use and may include short summaries, suspected causes, and suggested checks. Each response would include explicit references to the retrieved logbook entries, event records, archive segments, or documentation used to generate the answer, ensuring transparency and supporting validation in a safety-critical environment.

\subsection{Time-Correlated Archive Context}

To correlate events with machine parameter behavior, the framework aims to support time-aligned retrieval. Event time stamps would be used to retrieve archive variables within configurable pre- and post-event windows. Retrieved segments would be summarized using statistical descriptors such as extrema, mean values, and trends, and inserted into the language model context together with relevant logbook and event records.

\subsection{Current Limitations}
Current challenges include inconsistent naming conventions across subsystems, variability in logbook documentation quality, and imperfect time synchronization between data sources. Retrieval performance also depends on corpus coverage and update frequency. Proposed work would focus on improved subsystem tagging, richer metadata integration, and evaluation of retrieval performance and latency under realistic operator queries.
\section{SYSTEM ARCHITECTURE AND RAG PIPELINE}
The RAG framework is proposed to be implemented as a modular architecture separating data ingestion, semantic indexing, retrieval, and response generation. Figure~\ref{fig:rag_architecture} illustrates the overall system architecture. The design supports traceable responses, subsystem-aware filtering, and integration of structured and unstructured accelerator data.
The system consists of four main layers:

\textbf{Data Layer:} Electronic logbooks, interlock and machine protection records, archived process variables, and subsystem documentation.

\textbf{Indexing Layer:} Text chunking, metadata enrichment, embedding generation, and storage in a vector database.

\textbf{Retrieval Layer:} Hybrid semantic search combining similarity retrieval with metadata filtering and time constraints.

\textbf{Generation Layer:} Context-conditioned response generation using a large language model with enforced source traceability.

\subsection{Data Processing and Indexing}

Textual sources (logbooks, manuals, technical reports) are expected to be segmented into semantically coherent chunks and enriched with metadata fields including subsystem label, component identifier, time stamp (when available), and source type. Event-based records such as interlock triggers would be stored as structured entries containing event time, subsystem, and affected components.
Archived time-series process variables are planned to be accessed through a dedicated interface. For semantic integration, selected time windows would be summarized into compact statistical descriptors before being included in the language model context when required.
All textual and event records would be embedded using a domain-adapted embedding model and stored together with metadata in a vector database supporting similarity search and structured filtering.

\subsection{Query and Retrieval Pipeline}

When an operator submits a query, the system aims to execute the following workflow:
\begin{verbatim}
Input: User query Q
1. Normalize Q (acronyms, components, time hints)
2. Compute embedding v_Q
3. Retrieve top-k contexts from vector database
4. Apply metadata filtering (subsystem, time)
5. If event time detected:
      retrieve corresponding archive window
      compute statistical summary
6. Assemble prompt {Q, contexts, archive summary}
7. Generate response with LLM and source references

Output: Structured response with traceable sources
\end{verbatim}

This hybrid retrieval approach is expected to ensure that generated responses remain grounded in facility-specific operational data.





\bibliographystyle{unsrt}
\bibliography{WEP6008}

\end{document}